\documentclass[aps,prc,onecolumn,superscriptaddress,
showpacs,showkeys]{revtex4}
\usepackage{graphicx}
\usepackage{float} 
\usepackage{bm}
\usepackage{epsfig}
\usepackage{amsfonts}
\usepackage{amssymb,amscd}
\usepackage{xcolor}

\begin{document}
	
\title{Charm multiplicity distribution in high energy pp collisions with PYTHIA}

\author{Yuri N. {\sc Lima}}
\email{ylima@if.usp.br}
\affiliation{
Instituto de F\'{\i}sica - Universidade de S\~ao Paulo
Rua do Mat\~ao 1371 - CEP 05508-090\\
Cidade Universit\'aria, S\~ao Paulo - Brasil}

\author{Cristiane  {\sc Jahnke}}
\email{cjahnke@unicamp.br}
\affiliation{
Instituto de F\'{\i}sica Gleb Wataghin -  Universidade Estadual de Campinas
Rua S\'ergio Buarque de Holanda, 777 - CEP 13083-859\\
Campinas, S\~ao Paulo - Brasil}

\author{Marcelo G. {\sc Munhoz}}
\email{munhoz@if.usp.br}
\affiliation{
Instituto de F\'{\i}sica - Universidade de S\~ao Paulo
Rua do Mat\~ao 1371 - CEP 05508-090\\
Cidade Universit\'aria, S\~ao Paulo - Brasil}

\author{Fernando S. {\sc Navarra}}
\email{navarra@if.usp.br}
\affiliation{
Instituto de F\'{\i}sica - Universidade de S\~ao Paulo
Rua do Mat\~ao 1371 - CEP 05508-090\\
Cidade Universit\'aria, S\~ao Paulo - Brasil}
	
\begin{abstract}
With the growth of statistics in the future experiments at the LHC, the number of events with charm production 
will increase substantially. It may become possible to measure the multiplicity distribution of charm particles.
Using PYTHIA-8, we generated charm multiplicity distributions in  $pp$ collisions for different pseudorapidity ranges ($|\eta| < 0.5, 1.0, 2.0, 3.0$) and center-of-mass energies ($\sqrt{s}=$ 0.9, 2.36, 2.76, 7.0, 8.0, 13.0 TeV). We investigated the role played by multiple parton interactions and color reconnection. We compared the multiplicity distribution of D mesons with the charm quarks multiplicity distribution. We observe that the 
"quark-hadron" duality hypothesis is satisfied. With the obtained distributions we tested the validity of the Koba-Nielsen-Olesen scaling.  We also parameterized the charm distributions considering the Poisson and negative binomial distributions. 
\end{abstract}

\date{\today}
\maketitle

\section{Introduction}

In recent years, a large amount of data on multiplicity distributions, especially of charged particles, has been collected and made available by various experimental collaborations \cite{ALICE:2017pcy, ALICE:2025woy,
CMS:2010qvf, CMS:2018nhd, ATLAS:2010jvh, ATLAS:2016zba,ATLAS:2016zkp}. Several phenomenological studies  addressing these data are also available in the literature 
\cite{Duan:2025ngi,Islam:2025uns,Dokshitzer:2025fky,Dokshitzer:2025owq,Kulchitsky:2023fqd,Levin:2024wtl,
Grosse-Oetringhaus:2009eis,Martins-Fontes:2025iee,Martins-Fontes:2025xyq,Germano:2021brq,Germano:2024ier} 
. Particle multiplicity is a global observable that enables the characterization of events in colliding systems.  The multiplicity distribution can provide information about the dynamics of hadronic interactions at high energies and the formation of new states of matter. Analyzing this observable we can look for qualitative changes in the reaction dynamics.  

One of the expected changes is a transition between the soft and hard regimes of Quantum Chromodynamics (QCD). Another one is the onset of parton saturation and the formation of the Color Glass Condensate (CGC) 
\cite{Gelis:2010nm}. Finally,  in ultracentral  collisions we may have the formation of Quark-Gluon Plasma (QGP)\cite{Shuryak:2025byj}.   
In the absence of QGP, in high-energy proton-proton ($pp$) collisions, particles can be produced mainly in the following ways:
i) in an early stage of the collision, a perturbative parton cascade process occurs, governed by QCD evolution equations \cite{Kovchegov:2012mbw}, such as the  BFKL and the DGLAP equations, which determine how the parton distribution evolves with increasing collision energy and with the squared four-momentum of the probe; 
ii) during the collisions inelastic parton-parton collisions may produce new partons; 
iii) in the late stage of the  collision  partons hadronize and there is additional particle production. 
At increasing energies, we expect to observe the dominance of the first way and eventually the emergence of 
saturation.

The production of a heavy quark-antiquark pair comes from  hard parton-parton scatterings that occur at the 
early stages of high energy hadronic collisions. The production cross section can be calculated with perturbative QCD (pQCD). The production of multiple heavy quark pairs in the same event raises new questions concerning the 
nature of the emitting source. The charm multiplicity distribution can help in answering these questions and  improve our understanding about the production mechanism.

The increased luminosity and detector upgrades achieved during the Run-3 of the LHC, particularly in the ALICE
experiment, significantly improved the sensitivity to charm production in proton–proton collisions
\cite{ALICE:2022wwr}. These advances allow detailed studies of charm yields and correlations as a function of event activity, as well  as the extraction of moments of the charm multiplicity distribution 

In view of the forthcoming data, we would like to know what to expect. 
So far, to the best of our knowledge, there is no model for the multiplicity  distribution of charmed particles formed in $pp$ collisions. In this situation, a natural first step is to use the most successful model of 
particle production to obtain a baseline estimate. The most well known model is PYTHIA \cite{Bierlich:2022pfr,Sjostrand:2006za}.

In this work, we  use the PYTHIA-8 event simulator to  make the first prediction of charm multiplicity distributions, computed at different center-of-mass energies and different pseudorapidity windows. We 
compare these distributions with the charged particle distributions. In addition, we investigate the 
Koba-Nielsen-Olesen (KNO) scaling properties \cite{Koba:1972ng} 
of the obtained distributions and fit them to Poisson distributions, to single Negative Binomial Distribution (NBD), as well as to the weighted sum of two NBDs.

The text is organized as follows: in Section \ref{sec2} we briefly introduce the PYTHIA event generator;  in Section \ref{sec3} we  present the charm multiplicity distributions and compare them with the charged particle distributions;  in Section \ref{sec4} we discuss KNO scaling in the obtained distributions; in Section \ref{sec5} we present the parameterization of the distributions with Poisson and NBD fits; we conclude by summarizing the main results.

\section{PYTHIA event generator}
\label{sec2}

PYTHIA is an event generator designed to simulate particle collisions at high energies. An event is the outcome of a complete simulation of a  collision (from the initial instant of interaction to the final state with detectable particles). Each event yields a list of particles with their momenta, energies, masses, and generation history. PYTHIA simulates the interaction between two particles by considering a sequence of steps that model the main phenomena of high-energy particle physics: hard processes, parton showers from the initial and final states, multiple interactions of the partons, and hadronization and decays.

Proton-proton  collisions involve multiple perturbative scattering processes between the incident partons, which are implemented within the MultiParton Interactions (MPI) framework \cite{Sjostrand:1987su,Sjostrand:2004pf}. 
The perturbative scattering processes are accompanied by initial-state radiation and final-state radiation. Finally, PYTHIA uses the Lund string fragmentation model \cite{Andersson:1983ia} to implement hadronization, where the created partons and the beam remnants are connected by strings (color flux tubes). When the partons move away the strings break generating light quark-antiquark pairs.  When we consider Color Reconnection \cite{Argyropoulos:2014zoa}, the strings are reorganized, reducing their total length, allowing partons of different PIs to connect with each other. Because many particles are produced by string breaking, this reduction implies a reduction in the total multiplicity. Regarding uncertainties, PYTHIA generates Monte Carlo events based on stochastic models and does not automatically compute statistical errors. Because it generates event samples based on probabilistic distributions, it provides deterministic values based on the parameters and models configured by the user.

The results presented in this work are obtained from simulations using PYTHIA version 8.313. We generated 20 million events at different center-of-mass energies ($\sqrt{s}=$0.9, 2.36, 2.76, 7.0, 8.0, 13 TeV) and in different pseudorapidity windows ($|\eta| < $ 0.5, 1.0, 2.0, 3.0). The simulations presented in this work were performed considering all hard QCD processes. However, very similar results were observed when considering soft (low-energy) and hard (high-energy) QCD processes. For the charm  simulations, we considered the individual quark and antiquark counts without the effects of hadronization. In other words, we count the number of charm (and anticharm) quarks present in a stage immediately preceding the hadronization phase. For comparison, we also ``looked'' at the charm after the hadronization process, counting the number of $D$ mesons that form at this stage. For the simulations of $D$ mesons (stable and likely to reach the detector) we considered the sum of the production of well-established charmed $D$ mesons ($D^+$, $D^0$, $D^{*+}$ $D^{*0}$, $D^{*+}_2$, $D^{*0}_2$, $D^{+}_{s}$, $D^{*+}_{s}$, $D^{*+}_{s2}$) and their counterparts, disregarding the count of those that could originate from decays.

In PYTHIA the bulk of charm production takes place in multiple parton-parton scatterings, typically $g \, g \to c \, \bar{c}$.  Since these scatterings are independent from each other \cite{Sjostrand:2004pf},  the multiplicity distribution of charm pairs should be approximately a Poisson distribution, i.e., much narrower than the observed multiplicity distribution of charged particles. 

\section{Charm multiplicity distribution}
\label{sec3}

\begin{figure}
\includegraphics[page=1,width=0.49\textwidth]{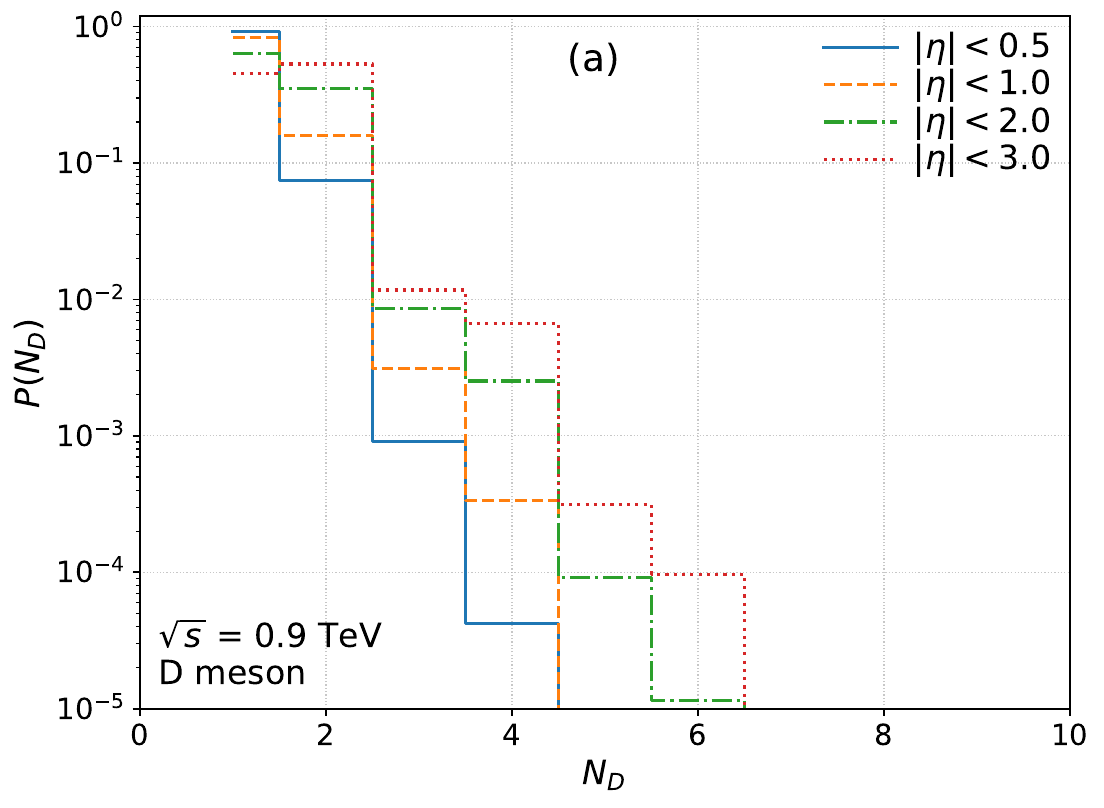}
\includegraphics[page=3,width=0.49\textwidth]{multiplicity_Dmeson.pdf}
\includegraphics[page=4,width=0.49\textwidth]{multiplicity_Dmeson.pdf}
\includegraphics[page=6,width=0.49\textwidth]{multiplicity_Dmeson.pdf}
\caption{$D$ meson (hadronic level) multiplicity distribution: comparison of different pseudorapidity range ($|\eta| <$ 0.5, 1.0, 2.0, 3.0).  We show results for a) $\sqrt{s} = 0.9$ TeV; b) $\sqrt{s}=  2.76$ TeV; c) 
$\sqrt{s}= 7$ TeV and d) $\sqrt{s}= 13$ TeV.}
\label{fig_Dmultiplicity}
\end{figure}

\begin{figure}
\includegraphics[page=1,width=0.49\textwidth]{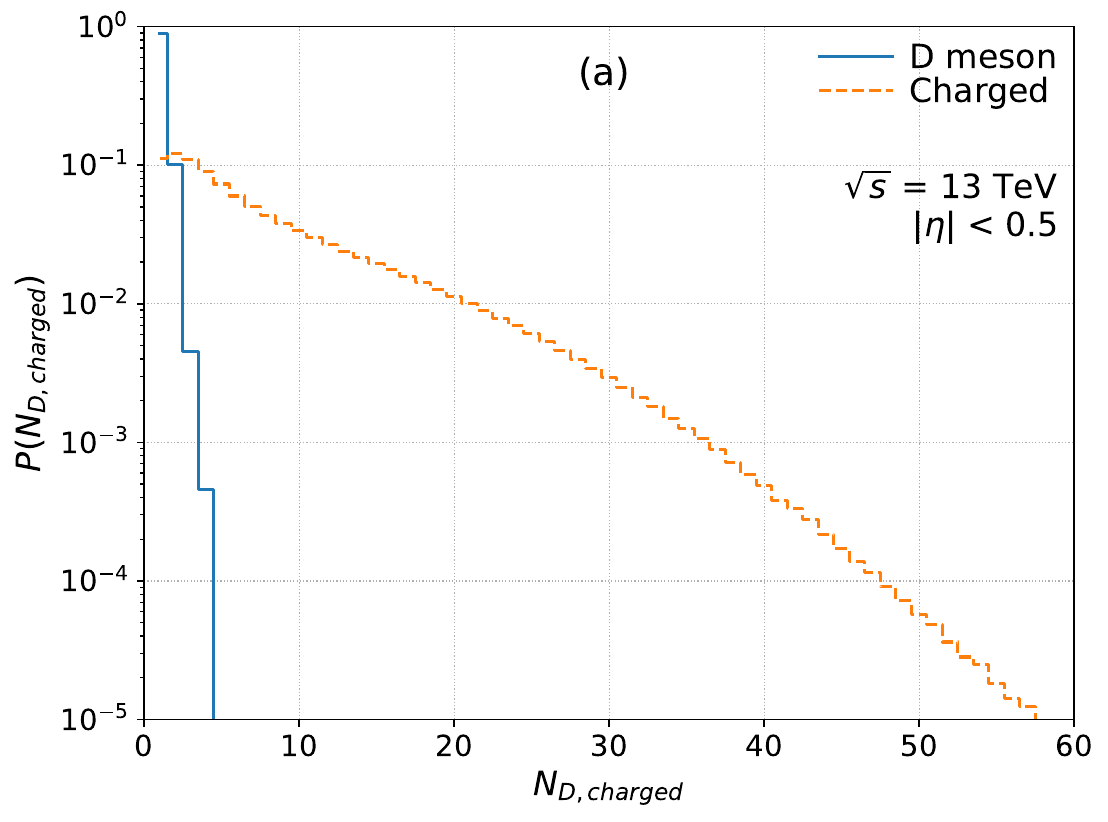}
\includegraphics[page=2,width=0.49\textwidth]{plot_charmchargedcomparison.pdf}
\caption{Comparison between $D$ meson and charged particle multiplicity distributions in $pp$ collisions at 13 TeV. 
Panel  a)  shows results for $|\eta|<$ 0.5  and panel b) for $|\eta|< $ 3.0).}
\label{fig_comparisoncharmcharged}
\end{figure}

In Fig. \ref{fig_Dmultiplicity} we present the $D$ meson multiplicity distributions for $pp$ collisions, where each panel shows results for different center-of-mass energies ($\sqrt{s}=$ 0.9, 2.76, 7.0, 13.0 TeV) with a comparison between several pseudorapidity ranges ($|\eta| <$ 0.5, 1.0, 2.0, 3.0). We can see that: the distributions reach a maximum at the beginning of the curve and then drop several orders of magnitude; the slope of the drop decreases slightly with increasing center-of-mass energy, and similarly with increasing pseudorapidity window; at small values of $N$ the curves approximately overlap; for large $N$ the difference between the curves becomes evident. As expected, the number of charm particles increases with increasing center-of-mass energy and with the pseudorapidity range considered. These observations are very similar to what one might find in an analysis of the multiplicity of charged particles. The difference is in the magnitude since charged particles are produced more abundantly than charm quarks, as shown in Fig. \ref{fig_comparisoncharmcharged}. 

In Fig. \ref{fig_comparisoncharmcharged} we show the comparison between charm and charged particle distributions measured in $pp$ collisions at 13 TeV. Panel (a) shows results for $|\eta| <$ 0.5 and panel (b) for $|\eta| <$ 3.0. As previously mentioned, charm quark multiplicities follow distributions that decrease sharply and are much narrower than those of charged particles. This happens because of the nature of the production process, i.e. the  multiple independent parton-parton scatterings mentioned above. Besides, charged particles are produced in large numbers whereas charm production is relatively rare and therefore the charm multiplicity spectrum is much narrower than that of charged particles. We can also see that the difference between the quantities increases as the pseudorapidity considered increases.

\begin{figure}
\includegraphics[page=1,width=0.49\textwidth]{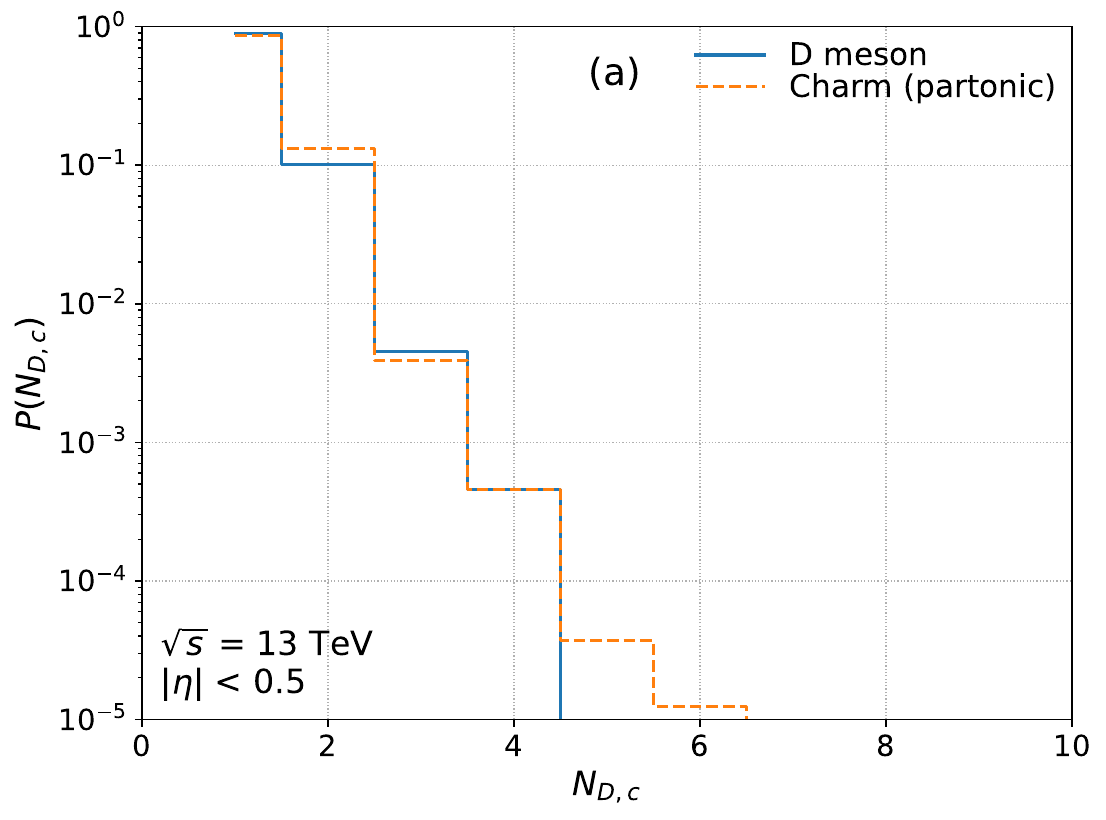}
\includegraphics[page=2,width=0.49\textwidth]{plot_charmcomparison.pdf}
\caption{Comparison between charm quark (parton level) and D meson (hadronic level) multiplicity distributions in 
$pp$ collisions at 13 TeV  for two pseudorapidity ranges a) $|\eta|<$ 0.5 and  b)  $|\eta|<$  3.0.}
\label{fig_comparisoncharmDmeson}
\end{figure}

The partonic level charm distribution projects mostly onto $D$ mesons. This implies that the shape of the partonic charm multiplicity distribution will be similar to that of D mesons, with an almost one-to-one conversion, as shown in Fig. \ref{fig_comparisoncharmDmeson}. According to the so-called parton-hadron duality hypothesis, the partonic and hadronic multiplicity distributions should be approximately the same. In fact, this was shown \cite{Duan:2025ngi}  to be the case for charged particles measured in jets. 

\begin{figure}
\includegraphics[page=1,width=0.49\textwidth]{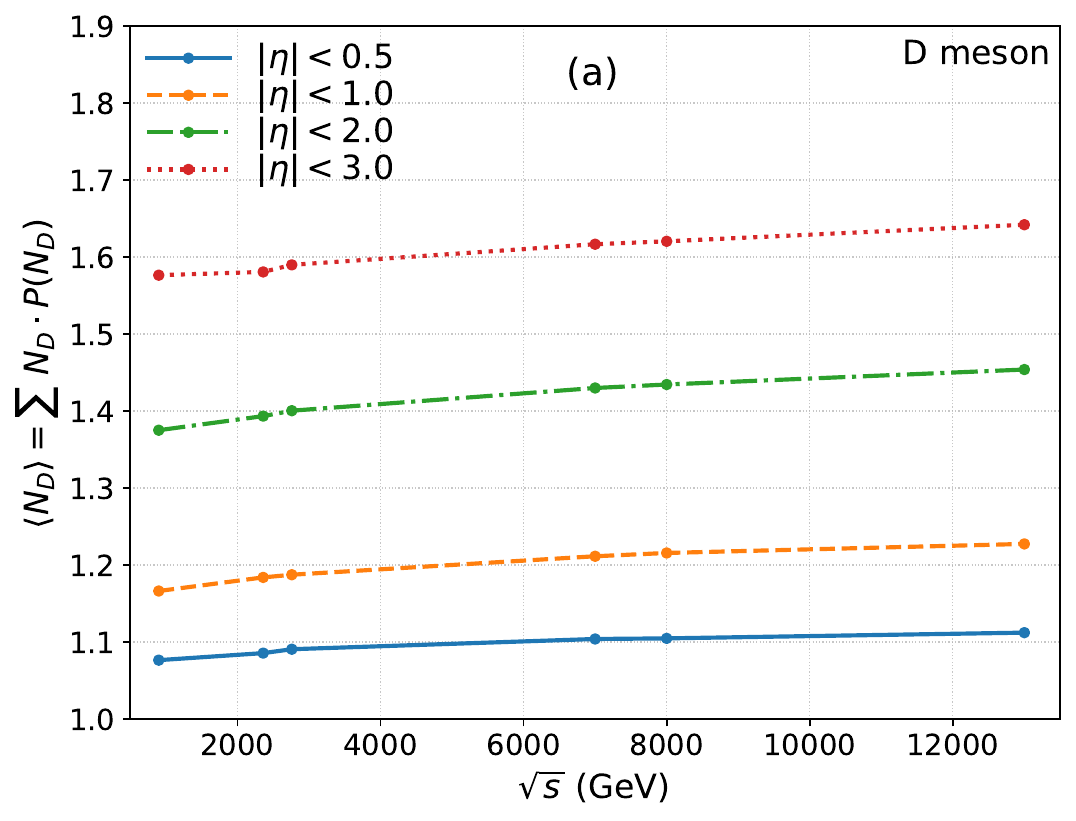}
\includegraphics[page=1,width=0.49\textwidth]{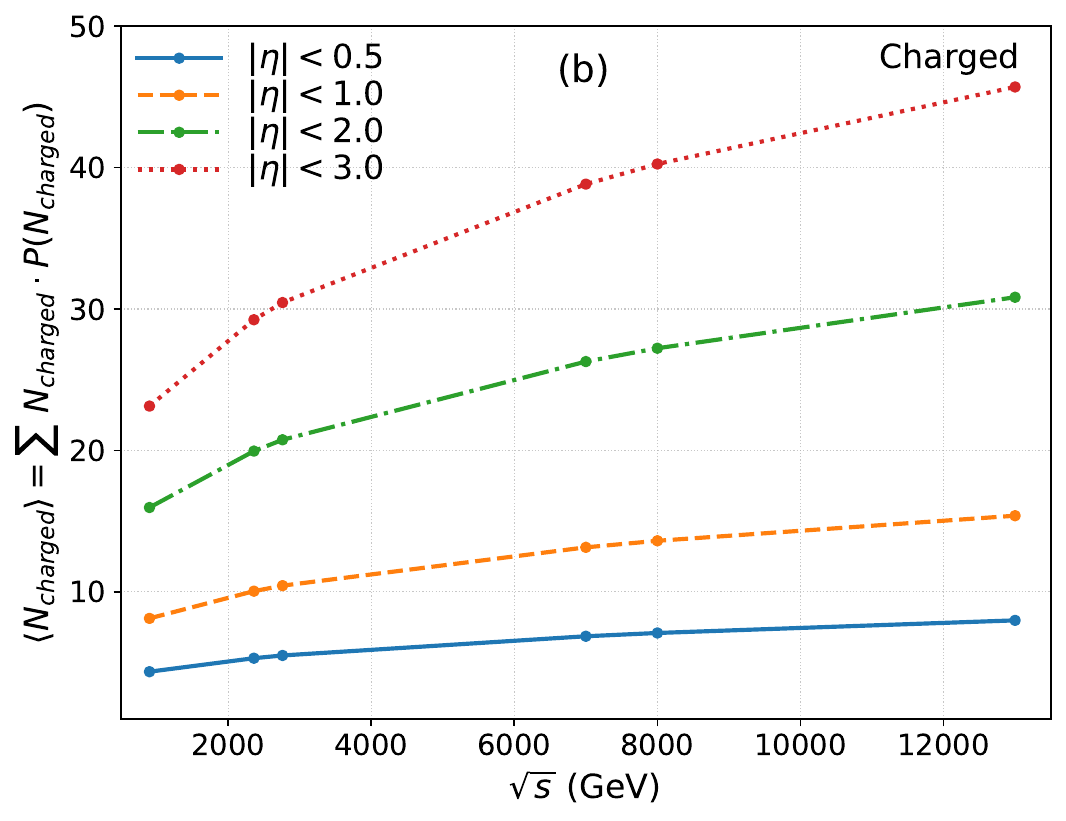}
\caption{a) Average D meson multiplicity as a function of the center-of-mass energy. b) 
Average charged particle multiplicity as a function of the center-of-mass energy.}
\label{fig_Naverage}
\end{figure}

Using the charm  multiplicity distribution, we can calculate the average multiplicity $\langle n_{D} \rangle$, 
which is shown in Fig. \ref{fig_Naverage}a as a function of the energy $\sqrt{s}$. An interesting feature of this figure is that $\langle n_{D} \rangle$ grows slowly with the energy and at the same rate for all the pseudorapidity intervals. In contrast,  the multiplicity of charged particles shown in Fig.  \ref{fig_Naverage}b, grows faster with
the energy. Moreover it also grows faster for the narrowest pseudorapidity window ($|\eta| < 0.5 $). We note that  
the charm production cross section grows faster than the total inelastic cross section. This means that, in PYTHIA, increasing the energy it rapidly becomes easier to produce {\it one} $c \bar{c}$ pair but the probability of producing two pairs (and  also three and so on) does not grow so fast.

\begin{figure}
\includegraphics[page=1,width=0.49\textwidth]{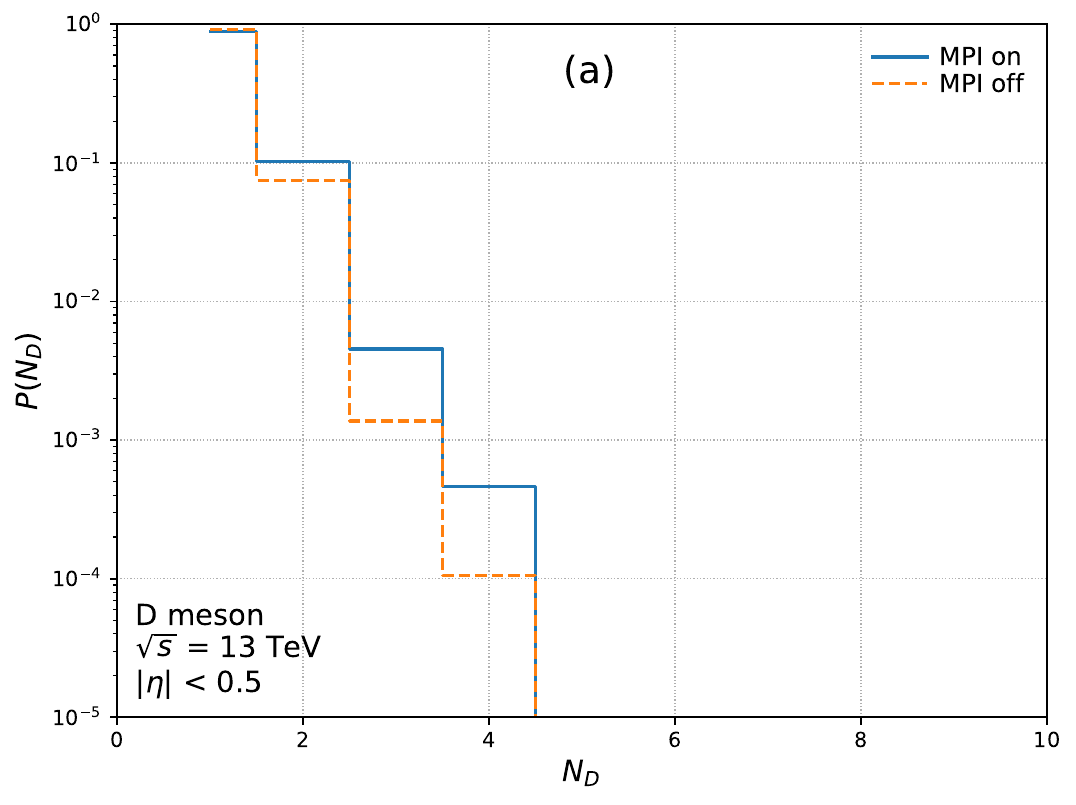}
\includegraphics[page=2,width=0.49\textwidth]{plot_MPIcomparison_Dmeson.pdf}
\caption{Effect of including multiple parton interactions (MPI) in  $pp$ collisions at $\sqrt{s} = 13$ TeV.
a) $|\eta| < 0.5$. b) $|\eta| < 3$.}
\label{fig_MPI}
\end{figure}

\begin{figure}
\includegraphics[page=1,width=0.49\textwidth]{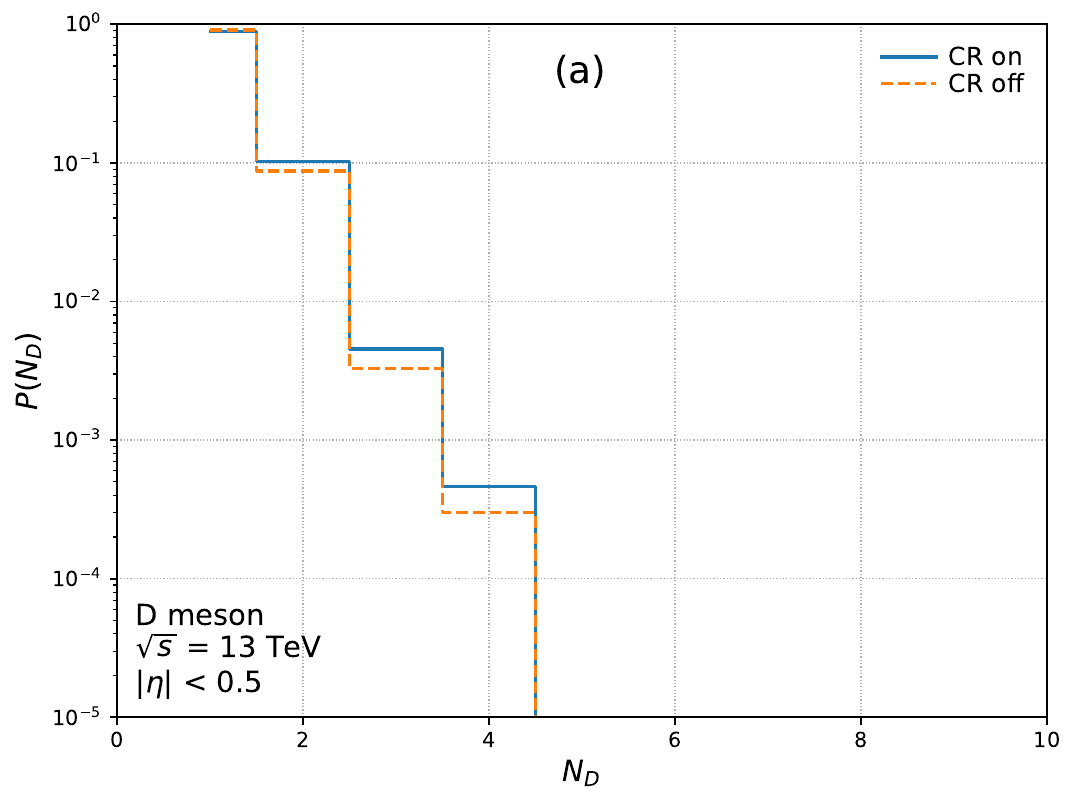}
\includegraphics[page=2,width=0.49\textwidth]{plot_CRcomparison_Dmeson.pdf}
\caption{Effect of including color reconnection (CR) in $pp$ collisions at $\sqrt{s} = 13$ TeV.
a) $|\eta| < 0.5$. b) $|\eta| < 3$.}
\label{fig_CR}
\end{figure}

In Fig. \ref{fig_MPI} we show the effect of multiple parton interactions (MPI).  As expected, switching MPI on leads to larger multiplicities. However, this effect is relatively small for $|\eta| < 0.5$ and is large for $|\eta| < 3$.
This means that small and large pseudorapidity windows have a somewhat different physics. 
In Fig. \ref{fig_CR} we show the effect of color reconnection (CR). We can see that this effect is very small for both cases $|\eta| < 0.5$ and $|\eta| < 3$.

\section{KNO scaling in charm multiplicity distributions}
\label{sec4}

The Koba-Nielsen-Olesen (KNO) scaling \cite{Koba:1972ng}, originally derived from the Feynman scaling, is a fundamental feature of multiplicity distributions, which suggests that at sufficiently high (asymptotic) energies the multiplicity exhibits universal scaling behavior. Thus, in a given collision, the probability $P(n)$ of producing $n$ particles can be written as
\begin{equation}
P_{n}(s) = \frac{1}{\langle n(s) \rangle} \, \Psi\left( \frac{n}{\langle n(s) \rangle} \right),
\end{equation}
where $\langle n(s) \rangle$ is the average particle multiplicity at a given center-of-mass value $\sqrt{s}$. From this, we expect the scaled distributions to become energy independent, i.e., the multiplicity distribution at different energies collapses into a universal curve. 

There are theoretical and phenomenological studies available in the literature that provide important insights into the behavior of the KNO scaling in hadronic collisions. For example, in Refs.  \cite{Dumitru:2012yr,Dumitru:2012tw}, the authors try to understand this phenomenon in terms of quark-gluon dynamics. They show that the KNO scaling appears  when the theory describing the color charge fluctuations is approximately Gaussian, when the relevant coupling constant is running and  when nonlinear saturation effects are present.  Another example can be found in Ref. \cite{Mahmoud:2022gzy}, where the authors systematically tested the KNO scaling in $pp$ collisions using PYTHIA simulations at different center-of-mass energies and pseudorapidity ranges. The authors reported that the KNO scaling is valid approximately for $|\eta|<$ 0.5, but significant violations appear as we consider larger pseudorapidity values, especially for large $N$, implying increasing fluctuations and the breakdown of universality in the particle production mechanism. More recently, in Refs. \cite{Martins-Fontes:2025iee,Martins-Fontes:2025xyq}, considering an explicit separation between soft and semi-hard processes in the $k_T$ factorization approach, the authors observed that in $pp$ collisions and within the central pseudorapidity region the KNO scaling is valid. Studies on the multiplicity of charged particles measured in $p - Pb$ can be found in Ref. \cite{ALICE:2022xip}, where the authors also address the existence of KNO scaling. They conclude that none of the considered models is able to satisfactorily describe the high multiplicity region of the data. However a different conclusion was reached in \cite{Terra:2025lys}.

\begin{figure}
\includegraphics[page=1,width=0.49\textwidth]{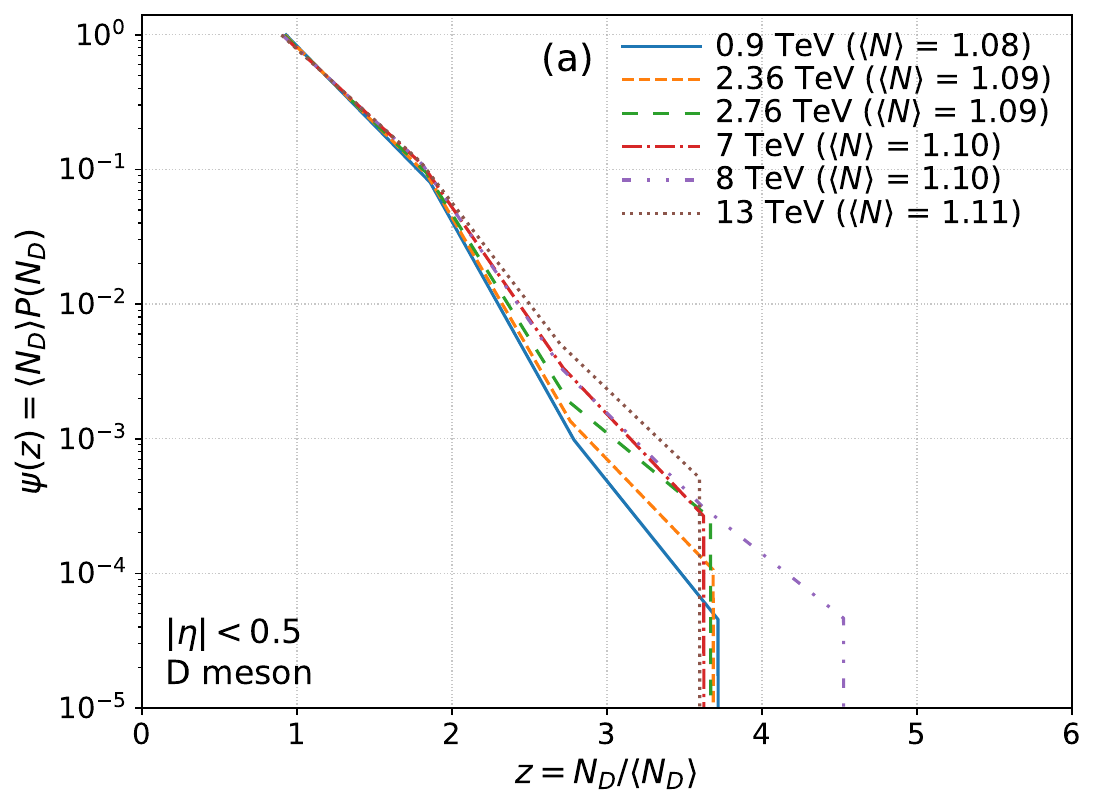}
\includegraphics[page=2,width=0.49\textwidth]{KNOscaling_Dmeson.pdf}
\includegraphics[page=3,width=0.49\textwidth]{KNOscaling_Dmeson.pdf}
\includegraphics[page=4,width=0.49\textwidth]{KNOscaling_Dmeson.pdf} 
\caption{$D$ meson multiplicity distribution in the KNO form for $pp$ collisions for different center-of-mass energies. a) $|\eta| <0.5$. b)  $|\eta| < 1$. c) $|\eta| < 2$. d) $|\eta| < 3$. }
\label{fig_KNODmeson}
\end{figure}

In Fig. \ref{fig_KNODmeson} we present the scaled charm multiplicity distributions for $pp$ collisions for  different pseudorapidity ranges. We can see that, in all panels, the curves  approximately overlap at small 
multiplicities but quickly begin to diverge. Furthermore, the tails increase with increasing energy. Stricly speaking  KNO scaling is violated in all pseudorapidity ranges.  However we might argue that for narrower pseudorapidity windows 
there is a residual scaling, whereas for wider windows the scaling is lost. This  behavior could be related to the
ocurrence of MPI at larger pseudorapidities.

\section{Parameterization with Poisson and NBD}
\label{sec5}

The Poisson distribution is a discrete probability distribution and works as a tool capable of modeling the probability of a number of events occurring, provided that these events occur at a constant average rate ($\lambda$) (and independently of the time since the last event). Therefore, here the events are independent, i.e., the 
occurrence of one event does not influence or affect the probability of occurrence of the next event. The Poisson 
distribution is given by:
\begin{equation}
P(n,\lambda)=\frac{\lambda^n e^{-\lambda}}{n!}
\end{equation}
where $n$ is the number of events and $\lambda$ is the average event rate. We may eliminate any zero contribution 
that could affect the Poisson distribution by using the zero-truncated Poisson distribution:
\begin{equation}
P(n,\lambda|n\geq 1,\lambda)= \frac{\lambda^n e^{-\lambda}}{1-e^{-\lambda}}.
\end{equation}
Another discrete probability distribution is the Negative Binomial Distribution (NBD), wich is used to describe the probability of observing a certain number of ``successes'' before a fixed number of ``failures''. In the context of particle physics, it is widely used to model the multiplicity distributions of charged particles produced in high-energy collisions. NBD is known to describe multiplicity distributions up to the energies achievable at the LHC. The fact that the NBD is successful in describing data is a strong indication that particle production is not an independent process, pointing to the presence of correlations between the final particles. The NBD distribution can be expressed as:
\begin{equation}
P(n,\langle n\rangle,k) = \frac{(n+k-1)!}{n!(k-1)!} (1-\langle n\rangle)^k \langle n\rangle^n
\label{NBDfunc}
\end{equation}
where $\langle n\rangle$ is the average of multiplicity and $k$ is a grouping parameter. The Poisson distribution is 
a particular case of NBD, when the parameter $k$ tends to infinity. 

Many experimental and theoretical studies show that in many cases, for high energies, it is necessary to consider the combination of two NBDs \cite{Giovannini:1998zb}, where each NBD corresponds to a different class of events: ``soft'' and ``semihard'' processes. This classification is based on different types of events and not on different particle production mechanisms within a single event. As a result, no interference term appears between the two components, and the multiplicity distribution can be written in the form 
\begin{equation}
P(n) = \lambda [\alpha_{soft} P_{NBD}(n,\langle n\rangle,k_1) + (1-\alpha_{soft})P_{NBD}(m,\langle m\rangle,k_2)]
\label{dnbd}
\end{equation}
However, unlike the case of charged particle production, the charm quark is produced in a small quantity, resulting in a narrow multiplicity distribution, as seen in the previous figures. Therefore, a double NBD is not expected to adequately describe the charm multiplicity distribution. The narrow shape of this distribution suggests a simple NBD distribution or even a Poisson form. In fact, when we try to fit the PYTHIA points with
(\ref{dnbd}) we find that $\alpha_{soft}=0$.

\begin{figure}
\includegraphics[page=6,width=0.49\textwidth]{fitfunctions_Dmeson.pdf}
\includegraphics[page=4,width=0.49\textwidth]{fitfunctions_Dmeson.pdf}
\includegraphics[page=3,width=0.49\textwidth]{fitfunctions_Dmeson.pdf}
\includegraphics[page=1,width=0.49\textwidth]{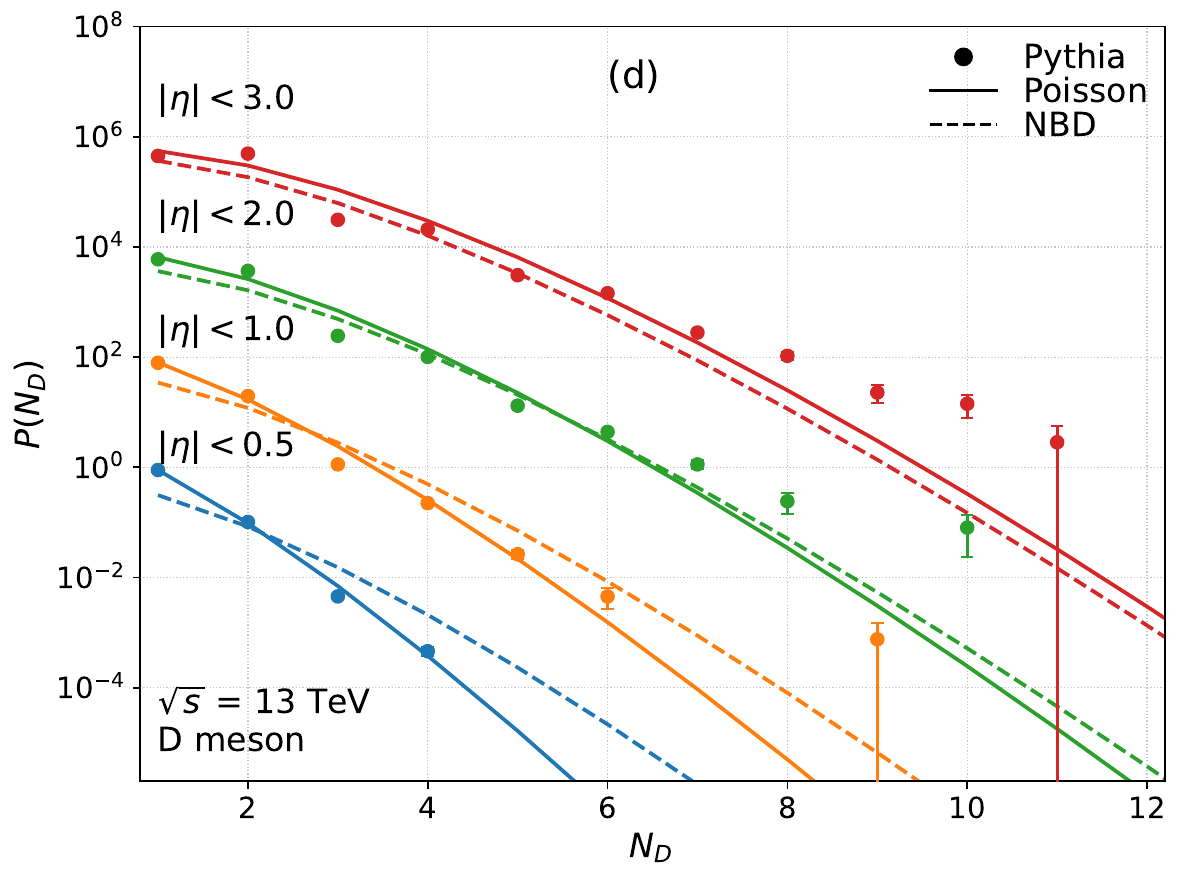}
\caption{Multiplicity distributions of $D$ meson fitted to Poisson and NBD distributions.
We show results for a) $\sqrt{s} = 0.9$ TeV; b) $\sqrt{s}=  2.76$ TeV; c) 
$\sqrt{s}= 7$ TeV and d) $\sqrt{s}= 13$ TeV.}
\label{fig_fitfunctionDmeson}
\end{figure}

In Fig. \ref{fig_fitfunctionDmeson} the charm meson multiplicity distributions, simulated with PYTHIA are compared with Poisson  and NBD fits for different pseudorapidity ranges ($|\eta|<$ 0.5, 1.0, 2.0, 3.0) with each panel for a distinct center-of-mass energy ($\sqrt{s}=$ 0.9, 2.76, 7.0, 13.0 TeV). We can see that, in all panels, for lower pseudorapidities the Poisson fit is closer to the points than the NBD fit; for higher pseudorapidities and lower energies the Poisson fit is above the simulated points and the NBD fits are closer to the points; for higher pseudorapidities and higher energies both fits are closer to the points (although the Poisson curve is higher than the NBD). As can be seen, neither of the fits reproduces all points; they always approximate one side or the other, but never the complete set of points.  This is due to the low statistics, i.e.,  the multiplicity has a low average, a low count, noisy statistics, which implies large fluctuations. It is also worth noting that using the double NBD would not solve this problem, since the first component of the double NBD is the dominant component, and it passes exactly over the projection of the NBD.

\section{Conclusion}

In this work we have used  the PYTHIA Monte Carlo event generator to study open charm production at the LHC. We focused
on the multiplicity distribution (MD) of charm quarks and charm mesons. Our main conclusions are:

\vskip0.2cm
\noindent
-The MD of charm is much narrower than the MD of charged particles and is close to a Poisson distribution. 
This is not surprising since multiple charm production in PYTHIA comes from independent multiple parton-parton scattering. 

\vskip0.2cm
\noindent
-The average charm multiplicity grows slower than the average charged particle multiplicity. This can be regarded 
as a threshold effect. 

\vskip0.2cm
\noindent
-The charm MD is not very different from the charm meson MD, showing that hadronization does not distort the charm quark MD. This finding gives support to the ``quark-hadron duality'' hypothesis. 

\vskip0.2cm
\noindent
-Going from  smalller to larger rapidity windows the MD becomes broader allowing for larger average multiplicities. However, for all the windows  $\langle n_{D} \rangle$ grows with $\sqrt{s}$ in the same way. 

\vskip0.2cm
\noindent
-Multiple parton interactions are not relevant at  narrow pseudorapidity intervals but play a significant role
for large pseudorapidity intervals. Color reconnection effects are irrelevant. 

\vskip0.2cm
\noindent
-Strictly speaking, we never observe KNO scaling. However, neglecting the large N tail of the MD, it is possible to argue that we observe some scaling for $|\eta| < 0.5$ and scaling violation for $|\eta| > 1.0$. This would indicate that at larger rapidities charm production follows a different dynamics, just as charged particle production. 

\vskip0.2cm
\noindent
-An attempt to fit the points with the  double negative binomial  (DNBD) leads to the conclusion that the fraction of semihard events is very large and the fraction of soft events is negligible. This brings us back 
to the single NBD fit which is reasonable, just as the Poisson fit. 

As mentioned in the introduction, the improvements achieved during the Run-3 of the LHC  allow detailed studies of charm yields. Nevertheless, a direct measurement of the full event-by-event charm multiplicity distribution remains challenging in Run-3 due to reconstruction efficiencies and large corrections. In this context, the results presented in this work provide a necessary baseline, establishing expectations for the shape, width, and scaling properties of charm multiplicity distributions. Looking ahead, the proposed ALICE 3 \cite{ALICE:2022wwr} at the HL-LHC is expected to enable a qualitative step forward, making direct measurements of charm multiplicity distributions at the hadronic level feasible. Such measurements will allow a direct confrontation between experimental data and the baseline predictions discussed here, including tests of Poisson-like behavior, Negative Binomial parameterizations, and KNO scaling in charm production.

\section{ACKNOWLEDGEMENTS}

We are grateful to Eliana Marroquin and André Veiga Giannini for rich and useful discussions. We also thank 
Jhoão Gabriel Arneiro for sharing with us his preliminary studies on this subject.
This study was supported, in part, by FAPESP (contract number 2024/17836-9), by CNPq and by INCT-FNA.



\begin{thebibliography}{99}

\bibitem{ALICE:2017pcy}
S.~Acharya \textit{et al.} [ALICE],
Eur. Phys. J. C \textbf{77},  852 (2017).

\bibitem{ALICE:2025woy}
S.~Acharya \textit{et al.} [ALICE],
Eur. Phys. J. C \textbf{85},  919 (2025).

\bibitem{CMS:2010qvf}
V.~Khachatryan \textit{et al.} [CMS],
JHEP \textbf{01}, 079 (2011).

\bibitem{CMS:2018nhd}
A.~M.~Sirunyan \textit{et al.} [CMS],
Eur. Phys. J. C \textbf{78},  697 (2018).

\bibitem{ATLAS:2010jvh}
G.~Aad \textit{et al.} [ATLAS],
New J. Phys. \textbf{13}, 053033 (2011).

\bibitem{ATLAS:2016zba}
M.~Aaboud \textit{et al.} [ATLAS],
Eur. Phys. J. C \textbf{76}, 502 (2016).

\bibitem{ATLAS:2016zkp}
G.~Aad \textit{et al.} [ATLAS],
Phys. Lett. B \textbf{758}, 67 (2016).

\bibitem{Duan:2025ngi}
X.~P.~Duan, L.~Chen, G.~L.~Ma, C.~A.~Salgado and B.~Wu,
Phys. Rev. D \textbf{112},  094022 (2025).

\bibitem{Islam:2025uns}
M.~S.~Islam and T.~Sinha,
Int. J. Mod. Phys. E \textbf{34},  2540001 (2025).

\bibitem{Dokshitzer:2025fky}
Y.~L.~Dokshitzer and B.~R.~Webber,
JHEP \textbf{10}, 114 (2025). 

\bibitem{Dokshitzer:2025owq}
Y.~L.~Dokshitzer and B.~R.~Webber,
JHEP \textbf{08}, 168 (2025). 

\bibitem{Kulchitsky:2023fqd}
Y.~A.~Kulchitsky and P.~Tsiareshka,
JHEP \textbf{10}, 111 (2023). 

\bibitem{Levin:2024wtl}
E.~Levin,
Phys. Rev. D \textbf{111},  016019 (2025). 

\bibitem{Grosse-Oetringhaus:2009eis}
J.~F.~Grosse-Oetringhaus and K.~Reygers,
J. Phys. G \textbf{37}, 083001 (2010). 

\bibitem{Martins-Fontes:2025iee}
H.~R.~Martins-Fontes and F.~S.~Navarra,
Phys. Rev. D \textbf{112},  094045 (2025).

\bibitem{Martins-Fontes:2025xyq}
H.~R.~Martins-Fontes and F.~S.~Navarra,
Physics \textbf{7}, 57 (2025). 

\bibitem{Germano:2021brq}
G.~R.~Germano and F.~S.~Navarra,
Phys. Rev. D \textbf{105},  014005 (2022). 

\bibitem{Germano:2024ier}
G.~R.~Germano, F.~S.~Navarra, G.~Wilk and Z.~Wlodarczyk,
Phys. Rev. D \textbf{110},  034026 (2024). 

\bibitem{Gelis:2010nm}
F.~Gelis, E.~Iancu, J.~Jalilian-Marian and R.~Venugopalan,
Ann. Rev. Nucl. Part. Sci. \textbf{60}, 463 (2010). 

\bibitem{Shuryak:2025byj} For a very recent review with a good list of references, see
E.~Shuryak,
[arXiv:2508.07985 [hep-ph]].

\bibitem{Kovchegov:2012mbw}
Y.~V.~Kovchegov and E.~Levin,
Camb. Monogr. Part. Phys. Nucl. Phys. Cosmol. \textbf{33}, 1-350 (2012)
Oxford University Press, 2013, ISBN 978-1-009-29144-6. 

\bibitem{ALICE:2022wwr}
 [ALICE],
``Letter of intent for ALICE 3: A next-generation heavy-ion experiment at the LHC,"
 [arXiv:2211.02491 [physics.ins-det]]. 

\bibitem{Bierlich:2022pfr}
C.~Bierlich, S.~Chakraborty, N.~Desai, L.~Gellersen, I.~Helenius, P.~Ilten, L.~L{\"o}nnblad, S.~Mrenna, S.~Prestel and C.~T.~Preuss, \textit{et al.}
SciPost Phys. Codeb. \textbf{2022}, 8 (2022). 

\bibitem{Sjostrand:2006za}
T.~Sjostrand, S.~Mrenna and P.~Z.~Skands,
JHEP \textbf{05}, 026 (2006). 

\bibitem{Koba:1972ng}
Z.~Koba, H.~B.~Nielsen and P.~Olesen,
Nucl. Phys. B \textbf{40}, 317  (1972).

\bibitem{Sjostrand:1987su}
T.~Sjostrand and M.~van Zijl,
Phys. Rev. D \textbf{36}, 2019 (1987).

\bibitem{Sjostrand:2004pf}
T.~Sjostrand and P.~Z.~Skands,
JHEP \textbf{03}, 053 (2004)

\bibitem{Andersson:1983ia}
B.~Andersson, G.~Gustafson, G.~Ingelman and T.~Sjostrand,
Phys. Rept. \textbf{97}, 31 (1983).

\bibitem{Argyropoulos:2014zoa}
S.~Argyropoulos and T.~Sj{\"o}strand,
JHEP \textbf{11}, 043 (2014).

\bibitem{Dumitru:2012yr}
A.~Dumitru and Y.~Nara,
Phys. Rev. C \textbf{85}, 034907 (2012). 

\bibitem{Dumitru:2012tw}
A.~Dumitru and E.~Petreska,
arXiv:1209.4105 

\bibitem{Mahmoud:2022gzy}
M.~A.~Mahmoud,
Particles \textbf{5}, 96 (2022).

\bibitem{ALICE:2022xip}
S.~Acharya \textit{et al.} [ALICE],
Phys. Lett. B \textbf{845}, 138110 (2023)
[erratum: Phys. Lett. B \textbf{853}, 138700 (2024)]. 

\bibitem{Terra:2025lys}
R.~Terra, A.~V.~Giannini and F.~S.~Navarra,
arXiv:2510.12561 

\bibitem{Giovannini:1998zb}
A.~Giovannini and R.~Ugoccioni,
Phys. Rev. D \textbf{59}, 094020 (1999).



\end{thebibliography}

\end{document}